\documentstyle[epsfig,aas2pp4]{article}

\slugcomment{Not to appear in Nonlearned J., 45.}

\def\Rsun{R_\odot}
\def\Msun{M_\odot}

\def\Mm{\, {\rm Mm}}

\def\ref{\par\noindent
        \hangindent=0.7 true cm
        \hangafter=1{}}
\def\arcsec{{\hbox{\rlap{\rlap{\tt\char"0D}\hbox{\thinspace\tt\char"0D}}
\kern-4.8pt\raise1pt\hbox{$\mit\mathchar"017F$}}}}

\def\note #1]{{\bf #1]}}
\def\figdir{fig}
\def\etal{{\it et al.}}

\begin{document}

\title{Accurate Determination \\
of the Solar Photospheric Radius}

\author{T. M. Brown}
\affil{High Altitude Observatory, National Center for 
Atmospheric Research,%
\footnote{The National Center for Atmospheric Research is Sponsored by
the National Science Foundation}\\
P.O. Box 3000, Boulder, CO 80307, USA}

\and

\author{J. Christensen-Dalsgaard}
\affil{Teoretisk Astrofysik Center, Danmarks Grundforskningsfond, and \\
Institut for Fysik og Astronomi, Aarhus Universitet, DK-8000 Aarhus C,
Denmark}

\begin{abstract}
The Solar Diameter Monitor measured the duration of solar meridian
transits during the 6 years 1981 to 1987, spanning the declining half
of solar cycle 21.
We have combined these photoelectric measurements with models of the
solar limb-darkening function, deriving a mean value for the solar
near-equatorial radius of 695.508 $\pm$ .026 Mm.
Annual averages of the radius are identical within the measurement
error of $\pm$ .037 Mm.
\end{abstract}


\section{Introduction}

The Sun is the only star for which reasonably precise values of
the mass, surface radius and luminosity are known.
The solar mass $\Msun$ is known from planetary motion, with
accuracy limited only by the uncertainty in
the gravitational constant $G$.
The solar radius can in principle be obtained from direct
optical measurement of the solar angular diameter, given
the very accurate determinations of the mean distance between
the Earth and the Sun.
In solar modeling, the value $\Rsun = 695.99 \Mm$
(Allen 1973) has been commonly used.
The models are calibrated to this photospheric radius,
in the present paper defined by the point in the atmosphere where 
the temperature equals the effective temperature,
by adjusting some measure of the convective efficacy,
such as the mixing length.

Recent accurate observations of solar f-mode frequencies
from the SOI/MDI instrument on the SOHO satellite
(e.g. Kosovichev {\etal} 1997) have raised some doubts
over this value of $\Rsun$.
The frequencies of these modes are predominantly determined 
by $G \Msun/\Rsun^3$.
By comparing the observed frequencies with frequencies of
solar models calibrated to $\Rsun = 695.99 \Mm$
Schou {\etal} (1997) and Antia (1998) concluded that
the actual solar radius was smaller by about $0.3 \Mm$
than the assumed radius of the model.
Other aspects of the modeling of the solar f modes may
affect their frequencies at this level
(e.g. Campbell \& Roberts 1989; Murawski \& Roberts 1993;
Ghosh, Antia \& Chitre 1995).
Thus it is obviously important to obtain independent 
verification of the proposed correction to the solar radius.

There are indeed significant uncertainties associated with
the currently adopted radius value.
These are related to the problem of the definition of
the solar limb adopted in the radius determinations,
and the reduction of the measured value to the photosphere.
It is not clear how the value quoted by Allen (1973) was obtained.
However, it appears that the more recent determinations,
which are generally consistent with Allen,
in most cases refer to the inflection point of the solar limb intensity.
According to solar atmospheric models this corresponds to 
a height of about $0.3 \Mm$ above the photosphere,
thus perhaps accounting for the radius correction inferred
from the f-mode frequencies.

The uncertainty in the precise definition of the measured
values of the solar radius highlights the need to
combine the observations with careful modeling of the
quantity that is observed.
Here we consider a long series of observations obtained
with the High Altitude Observatory's 
Solar Diameter Monitor (Brown {\etal} 1982).
This is based on a definition of the solar limb which
minimizes the effect of seeing (Hill, Stebbins \& Oleson 1975).
By combining daily data obtained over more than 6 years,
extending between solar maximum and solar minimum,
the possible effects of solar activity can be checked.
The analysis of the data is carried out by means of
a model of the solar limb intensity, following 
as closely as possible the
actual procedure used in the reduction of the data
and testing for the effects of seeing.
In this way we have eliminated several of the uncertainties
affecting earlier determinations to arrive at what we
believe to be an accurate measure of the solar photospheric radius.

\section{Observations}

The Diameter Monitor instrument and its associated observing procedures
were described in detail by Brown {\etal} (1982);
it operated between August 1981 and December 1987.
It consisted of a meridian-transit telescope arranged to
allow the solar image to drift across a fixed detector package each
day at local noon.
A filter system confined the bandpass of the observed light to a 10 nm
band near 800 nm.
The horizontal (east/west in the sky) diameter
was obtained by timing the passage of the solar limbs across
each of two linear detector arrays aligned end-to-end.
Each detector pixel subtended 1{\arcsec} in the sky in the direction
of the apparent solar motion and 80{\arcsec} in the perpendicular direction.
In addition, the vertical (north/south in the sky)  diameter
was measured, although less precisely.
For the purposes of
this paper we shall therefore consider only the horizontal diameter.

An automated guiding system assured that the solar disk transited the
detector arrays centrally, so that a true diameter was measured.
Each readout of a detector array yielded a sample of the solar limb-darkening
function;
by reading the detectors at a 32~Hz rate, the instrument obtained
samples at intervals comparable to the seeing-change time.
The instrument
applied  a real-time edge-finding algorithm, and stored the
resulting edge positions.
This process was performed for the transits of
both west and east solar limbs, so a transit duration could be measured.
Ancillary quantities were also measured each day,
including seeing and scattered-light parameters.

The edge-finding algorithm used was the 
Finite Fourier Transform Definition (FFTD)
described by Hill {\etal} (1975).
The procedure involves forming the convolution of the observed 
limb-darkening function with a set of weights that are nonzero only within
a certain window of width $a$.
The edge was then defined to be the position of the center of the window
for which the convolution crossed zero.

The FFTD has two important features.
First, by a suitable choice of weights one can eliminate the first-order
sensitivity of edge position to seeing, for some chosen width of the seeing
point-spread function.
Daytime seeing is commonly both poor and variable,
and inflection-point definitions of the limb
position are highly sensitive to this variability.
Second, the FFTD depends on a free parameter, namely the window width $a$.
Applying the FFTD to the solar limb-darking function yields diameters $D(a)$
approximately equal to
$$
D(a) \ = \ D_0 \ - \ \alpha a \ ,
\eqno (1)
$$
where $D_0$ is the true angular distance between the nearly 
discontinuous intensity jumps
at east and west limbs,
and $\alpha$ is proportional to the intensity gradient
in the last 25{\arcsec} inside the limb,
and hence to the vertical temperature gradient in the photosphere
above an optical depth of roughly 0.2. 
By measuring the diameter over a range of $a$
and extrapolating the results to zero window width,
one can obtain a value that is largely independent of changes in
the slope of the limb-darkening function.
Since this slope proves to vary significantly during the solar
cycle, using multiple window widths is necessary to measure
the slope and remove its
effect from the raw diameter measurements.

\begin{figure}[hbt]
\epsfxsize=8.0cm \epsfbox{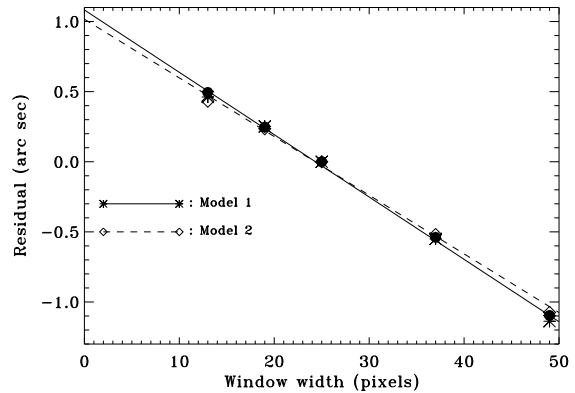}
\caption[]{
Observed and
inferred diameters ({\it i.e.,} separations between limb positions), 
as a function of window width.
Observed values are shown as filled circles, calculated values
for Model 1 as stars and Model 2 as diamonds.
Seeing with a FWHM of $6 \arcsec$ 
was assumed.
Also shown are the results of linear least-squares fits
to the limb positions, defining the extrapolation to zero window width.}
\end{figure}

Data were taken with the Diameter Monitor on any day for
which a successful observation seemed possible;
many of the observations were therefore corrupted by clouds or
(less often) instrument failures.
Accordingly, we used several selection criteria to choose the observations
to be used for the analysis,
in the end retaining 550 of the original 986 daily measurements,
each made with 5 different window widths.
Finally, we corrected the diameter values for
several geometrical sources of systematic error, and
projected all measurements to
a standard seeing width of 6{\arcsec}, which is the most frequently
observed value.

Figure 1 shows the unweighted average of these diameter values,
reduced to a Sun-Earth distance of 1 AU and plotted against the FFTD window
width.
Also shown are the results of applying the FFTD to the seeing-blurred
limb-darkening functions derived from two different
model solar atmospheres (see below).
The agreement between theory and observation is satisfactory, although
residual differences affect the necessary extrapolation to zero $a$,
and are a significant source of systematic error.

Figure 2 shows the measured time series of diameter measurements $D_0$
(projected to $a = 0$ by linear extrapolation) and of $\alpha$,
averaged over Carrington rotation periods of 27.275 days.
The error bars are standard deviations of the mean for each
rotation, estimated from the dispersion of the daily measurements
within that rotation.
The scatter among the daily diameter measurements is about 0.4{\arcsec}
rms, and arises mostly from time-dependent motions of the solar image
related to atmospheric seeing.
The diameter was essentially constant throughout the 6.3-year
observing period, aside from a possible but poorly-sampled upturn of
about 0.1{\arcsec} during 1987.
The limb-darkening function slope, however, varied with a time scale
of a year or more over a total range of about 2.5\%,
being steeper at solar minimum than at maximum.

We estimate $D_0$
as the unweighted mean of the 550 daily values, with
a random error equal to the standard deviation of the mean.
These values are
$$
D_0 \ = \ 1919.359 \arcsec  \pm 0.018 \arcsec \ \ .
\eqno (2)
$$

\begin{figure}[hbt]
\epsfxsize=8.0cm \epsfbox{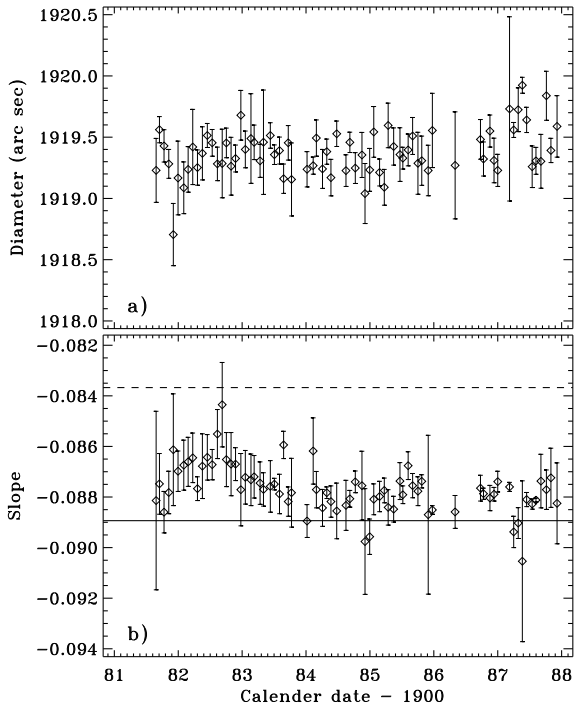}
\caption[]{
(a) Diameters extrapolated to zero window width,
averaged over intervals of one Carrington rotation,
shown as a function of calendar date.
(b) The limb-slope parameter $\alpha$,
averaged in the same fashion as in (a).
The solid and dashed horizontal lines indicate the slopes found for
Models 1 and 2.
}
\end{figure}

\section{Modeling}

As mentioned in the Introduction, the quantity to be measured is the
radius of the surface at which the local temperature equals the solar
effective temperature.
Evidently, this radius is related to $D_0$ in Eq. (2), 
but the relationship is not
a simple one;
it depends upon the radiation transfer in the outer solar atmosphere,
and upon the behavior of the FFTD limb definition.
To infer the correct radius from the observations, one must 
use a physically-based model of the solar atmosphere to calculate the
emergent intensity as a function of distance from the center
of the solar disk, 
and then compute the location on this brightness profile that would
be identified as the edge by the FFTD.
We calculated the limb intensity by integrating the equation
of transfer along rays through an assumed spherically symmetrical
solar atmosphere.
Since the observations were carried out in a relatively narrow
wavelength region around 800 nm, we considered simply the
monochromatic continuum intensity at this wavelength.

We have used two models of the solar atmosphere.
Model 1 was kindly computed
by Rodney Medupe with the ATLAS9 code (Kurucz 1993).
Model 2 was obtained
as an average of a hydrodynamical simulation of convection
in the upper part of the solar convection zone and lower
atmosphere, as described by, e.g., Stein \& Nordlund (1989)
but with updated physics (Trampedach 1997);
the average was performed at constant monochromatic optical
depth, at 800 nm.
The opacity was computed from the ATLAS data in both cases.
For Model 1 the source function $S_\lambda$ was obtained
from the ATLAS code, and hence allowed for mild departure from LTE.
For Model 2 LTE was assumed, so that $S_\lambda = B_\lambda$,
the Planck function.

To simulate effects of seeing, we convolved the intensity with
a gaussian, with full width at half maximum (FWHM)
specified in arc~sec and converted to linear distance at 1 AU.
After this convolution, we integrated the intensities over
pixels corresponding to $1 \arcsec$ at 1~AU, to match the
observed intensities.
We folded the pixel-weighted intensities with the FFTD weights
over the five different windows described in section 2,
and carried out the subsequent analysis to determine the limb position,
through extrapolation to zero window width, 
as for the observations.

The results of applying the observational procedure to the 
computed pixel-averaged intensities for Models 1 and 2,
assuming $6 \arcsec$ seeing
are shown in Fig. 1.
The observed variation of diameter with window
width and that calculated from the models agree well,
especially for Model 1.
Of particular interest are the extrapolations to zero window width,
corresponding to the observed limb position measured relative to 
the nominal photosphere of the models;
we obtain $0.47950 \; \Mm$ and $0.51634 \; \Mm$
for Models 1 and 2.

We have tested the sensitivity of the results to various assumptions
in the calculation. 
Replacing the true source function $S_\lambda$ by $B_\lambda$ for
Model 1 changes the limb position by much less than $0.001$ Mm;
thus the assumption of LTE in Model 2 is not a significant source of error.
Using an assumed seeing of less than 6{\arcsec} changes the limb position
by less than $0.01 \Mm$,
confirming the insensitivity of the FFTD to effects
of seeing.
However, the difference between the two models obviously remains
a source of some concern.

\section{Results and discussion}

We adopted the modified IAU (1976) value of 1.4959787066 $\times 10^5$ Mm
(Astronomical Almanac, 1997) for the astronomical unit, 
and adjusted this value by
-4.678 Mm to account for the mean displacement between the telescope's
noontime location and the Earth's center, and by +0.449 Mm
for the displacement of the Sun's center relative to the barycenter
of the Earth-Sun system.
This distance, combined with $D_0$ from Eq. (2), 
yields the Sun's apparent radius.
Applying the model corrections described in the last section,
we obtain
$$
\Rsun = (695.5260 \pm 0.0065 ) \Mm\qquad \hbox{\rm for Model 1}
$$
$$
\Rsun = (695.4892 \pm 0.0065 ) \Mm\qquad \hbox{\rm for Model 2} 
$$

We estimate the modeling errors to be $1/\sqrt 2$ of the difference
between these estimates, or about 0.020 Mm.
Based on the uncertainties in the geometric corrections that were
made to the measured radius, we estimate the systematic errors in the
measured value to be 0.015 Mm, or about twice as large as the random
errors.
Averaging our results for Models 1 and 2, and adding the various error
sources in quadrature, we arrive at our final estimate of
$$
\Rsun = (695.508 \pm 0.026 ) \Mm
$$

The inferred solar photospheric radius is smaller by about $0.5 \Mm$
than the normally used value of $695.99 \Mm$ (Allen 1976).
A review of recent observations was given by Schou {\etal} (1997),
concluding that these were consistent with an angular diameter
of $1919.26\arcsec \pm 0.2\arcsec$, corresponding to Allen's value
of $\Rsun$.
This is also consistent with the observed value obtained here (cf. eq. 2).
However, it appears that the observations considered by Schou {\etal}
refer to the inflection point of intensity (or, in one case, to 
an FFTD determination) and hence do not contain the correction to
photospheric radius.
Such a correction, taking into account the observational characteristics,
is an essential part of the radius determination.

Some confirmation of the reliability of the modeling comes from
the comparison in Fig. 2
of computed and observed slopes of the limb position
as function of the scan widths.
Nevertheless, it is striking that, as indicated by the difference
between Models 1 and 2, the major uncertainty in $\Rsun$ appears
to come from the modeling.
Indeed, it is evident that the real solar atmosphere is substantially
more complicated than the one-dimensional model resulting
from the ATLAS code or the mean model obtained from the hydrodynamical
simulations.
A more accurate determination of the radius correction
can probably be obtained from a
detailed calculation of the limb intensity,
taking into account the inhomogeneous nature of the relevant
layers, on the basis of the simulations.
Such an investigation is beyond the scope of the present paper, however.

We find no significant variation in the observed diameter
during the observation period (cf. Fig. 2); 
annual averages of the radius for the years 1981 to 1987
all agree within their measurement errors of $\pm .037$ Mm.
These limits are substantially smaller than
diameter changes reported previously for the same interval of time
(e.g. Ulrich \& Bertello 1995, Laclare {\it et al.} 1996),
but are in agreement with measurements by Wittman (1997).
On the other hand, the limb-position slope shows fairly substantial
variations.
We also note that during solar maximum, 
the daily slope values tended to be highly variable as well as small
in magnitude;
this suggests that the long-term variation may result from localized
activity-dependent features such as faculae.
It is plausible that the previously inferred variations in solar diameter
with solar activity is in fact a reflection of such variations
in the limb-darkening slope.

It is interesting that the value of $\Rsun$ obtained here is
somewhat smaller than that inferred from the solar f-mode frequencies,
indicating additional contributions to the differences between
the observed and model values of these frequencies.
This issue, and the
effects of the reduction of the model radius on the helioseismically
determined structure of the solar interior will be considered elsewhere.
We note, however, that Antia (1998) and Schou {\etal} (1997) found
significant effects on the helioseismically 
inferred sound speed from corresponding
radius changes.

\section*{Acknowledgements}

We are grateful to {\AA}. Nordlund, R. F. Stein and R. Trampedach
for providing the averaged hydrodynamical model, to R. Medupe for
the ATLAS model and to M. Kn\"olker for help with the opacity calculations.
This work was supported in part by the Danish National Research
Foundation through its establishment of the Theoretical Astrophysics Center.


\begin{thebibliography}{}

\bibitem[]{}
Allen, C. W., 1973.
{\it Astrophysical Quantities}, 3rd edition, p. 169
Athlone Press, London.
\bibitem[]{}
Antia, H. M., 1998.
A\&A, 330, 336
\bibitem[]{}
Astronomical Almanac, 1997.
Washington, U.S. Government Printing Office, p. K6.
\bibitem[]{}
Brown, T. M., Elmore, D. F., Lacey, L. \& Hull, H., 1982.
Appl. Optics, 21, 3588
\bibitem[]{}
Campbell, W. R. \& Roberts, B., 1989.
ApJ, 
338, 538
\bibitem[]{}
Ghosh, P., Antia, H. M. \& Chitre, S. M., 1995.
ApJ, 451, 851
\bibitem[]{}
Hill, H. A., Stebbins, R. T. \& Oleson, J. R., 1975.
ApJ, 200, 484
\bibitem[]{}
Kosovichev, A. G., Schou, J., Scherrer, P. H., {\etal}, 1997.
Solar Phys., 170, 43
\bibitem[]{}
Kurucz, R. L., 1993.
In {\it Peculiar versus normal phenomena in A-type and related stars},
Dworetsky, M. M., Castelli, F. \& Faraggiana, R. (eds),
ASP Conf. Ser., 44, 87
\bibitem[]{}
Laclare, F., Delmas, C., Coin, J.P., \& Irbah, A. 1996.
Solar Phys. 166, 211
\bibitem[]{}
Murawski, K. \& Roberts, B., 1993.
A\&A, 272, 595
\bibitem[]{}
Schou, J., Kosovichev, A. G., Goode, P. R. \& Dziembowski, W. A., 1997.
ApJ, 489, L197
\bibitem[]{}
Stein, R. F. \& Nordlund, {\AA}., 1989.
ApJ, 
342, L95
\bibitem[]{}
Trampedach, R., 1997.
{\it Convection in stellar atmospheres},
MSc Thesis, Aarhus University.
\bibitem[]{}
Ulrich, R.K. \& Bertello, L. 1995.
Nature 377, 214
\bibitem[]{}
Wittman, A.D. 1997.
Solar Phys. 171, 231

\end{thebibliography}
\end{document}